\documentclass[%
 reprint,
 superscriptaddress,
 amsmath,amssymb,
 aps,
]{revtex4-1}

\usepackage{graphicx}
\usepackage{dcolumn}
\usepackage{color}
\usepackage{bm}
\usepackage{hyperref}

\begin{document}

\preprint{APS/123-QED}

\title{An {\it ab initio} perspective on scanning tunneling microscopy measurements of the tunable Kondo resonance of the TbPc$_2$ molecule on a gold substrate}

\author{Ivan I. Vrubel}
\affiliation{Skolkovo Institute of Science and Technology, Moscow 121205, Russia}

\author{Anastasiia A. Pervishko}
\affiliation{Department of Physics and Astronomy, Uppsala University, Box 516, SE-751 20 Uppsala, Sweden}
\affiliation{ITMO University, Saint Petersburg 197101, Russia}

\author{Heike Herper}
\affiliation{Department of Physics and Astronomy, Uppsala University, Box 516, SE-751 20 Uppsala, Sweden}

\author{Barbara Brena}
\affiliation{Department of Physics and Astronomy, Uppsala University, Box 516, SE-751 20 Uppsala, Sweden}

\author{Olle Eriksson}
\affiliation{Department of Physics and Astronomy, Uppsala University, Box 516, SE-751 20 Uppsala, Sweden}
\affiliation{School of Science and Technology, \"Orebro University, SE-701 82 \"Orebro, Sweden}

\author{Dmitry Yudin}
\affiliation{Skolkovo Institute of Science and Technology, Moscow 121205, Russia}
\affiliation{Department of Physics and Astronomy, Uppsala University, Box 516, SE-751 20 Uppsala, Sweden}

\date{\today}

\begin{abstract}

With recent advances in the areas of nanostructure fabrication and molecular spintronics the idea of using single molecule magnets as building blocks for the next generation electronic devices becomes viable. A particular example represents a metal-organic complex in which organic ligands surround a rare-earth element or transition metal. Recently, it was explicitly shown that the relative position of the ligands with respect to each other can be reversibly changed by the external voltage without any need of the chemical modification of the sample. This opens a way of the electrical tuning of the Kondo effect in such metal-organic complexes. In this work, we present a detailed and systematic analysis of this effect in TbPc$_2$ from an {\it ab initio} perspective and compare the obtained results with the existing experimental data.

\end{abstract}

\maketitle

\section{Introduction}

Recently, enormous progress in fabrication of nanostructures brought attention to the field of metal-organic molecules. This interest is mainly motivated by the possibility of using these complexes for spintronics \cite{Rocha2005,Szulczewski2009}. The presence of the localized metallic shells in these compounds may lead to the appearance of the pronounced phenomena related to the Kondo effect \cite{Kondo1964}, which can be addressed in a highly controllable manner with state-of-the-art experimental methods \cite{Zhao2005}. A prototype Kondo system could be a quantum dot which is weakly coupled to two metallic electrodes \cite{Gordon1998,Cronenwett1998,Kouwenhoven2001}. An unpaired electron of the quantum dot represents a localized spin state which is coupled to the Fermi sea of conduction electrons in the leads via an exchange type of interaction. Below a characteristic Kondo temperature the localized spin and the spins of the conduction electrons form a singlet state. This Kondo state results in a resonant level at the Fermi energy of the leads, the so-called Kondo resonance. This resonance facilitates the tunneling of conduction electrons and leads to the appearance of the Kondo plateau in the conductance of the system as function of the gate voltage at low temperatures and zero bias voltage \cite{Ng1988,Glazman1988}.

The Kondo effect \cite{Kondo1964} has been observed in a variety of systems, e.g., carbon nanotubes \cite{Nygard2000}, semiconductor nanowires \cite{Jespersen2006}, divanadium molecules \cite{Liang2002}, fullerenes \cite{Park2002,Parks2010}, and complex organic structures, including single molecule magnets (SMMs) in the form of double-decker complexes with a 4$f$-like moment \cite{Ishikawa03,Zhao2005,Romeike2006,Vitali2008,Zhang09,Fu2012,Warner16}. The latter include, e.g., copper phthalocyanine (CuPc) \cite{Cinchetti2009} and CoPc molecules deposited on a nonmagnetic Au(111) substrate \cite{Zhao2005} and Co islands, where spin imaging of the spin of the central Co atom was reported in scanning tunneling microscopy (STM) measurements \cite{Iacovita2008,Atodiresei2010}. An important property of the Kondo effect in metal-organic complexes is the possibility of its electric tunability, as was experimentally demonstrated in Ref.~\onlinecite{Komeda2011}. Kondo-type features were identified in STM images of a double-decker molecular complex containing Tb, a rare-earth element. More specifically, the molecular system was bis-(phthalocyaninato)terbium (III), or simply TbPc$_2$, deposited on top of the Au(111) substrate, and a Kondo temperature of 31 K was inferred from these experiments. 

\begin{figure}[h!]
\includegraphics[scale=0.2,clip]{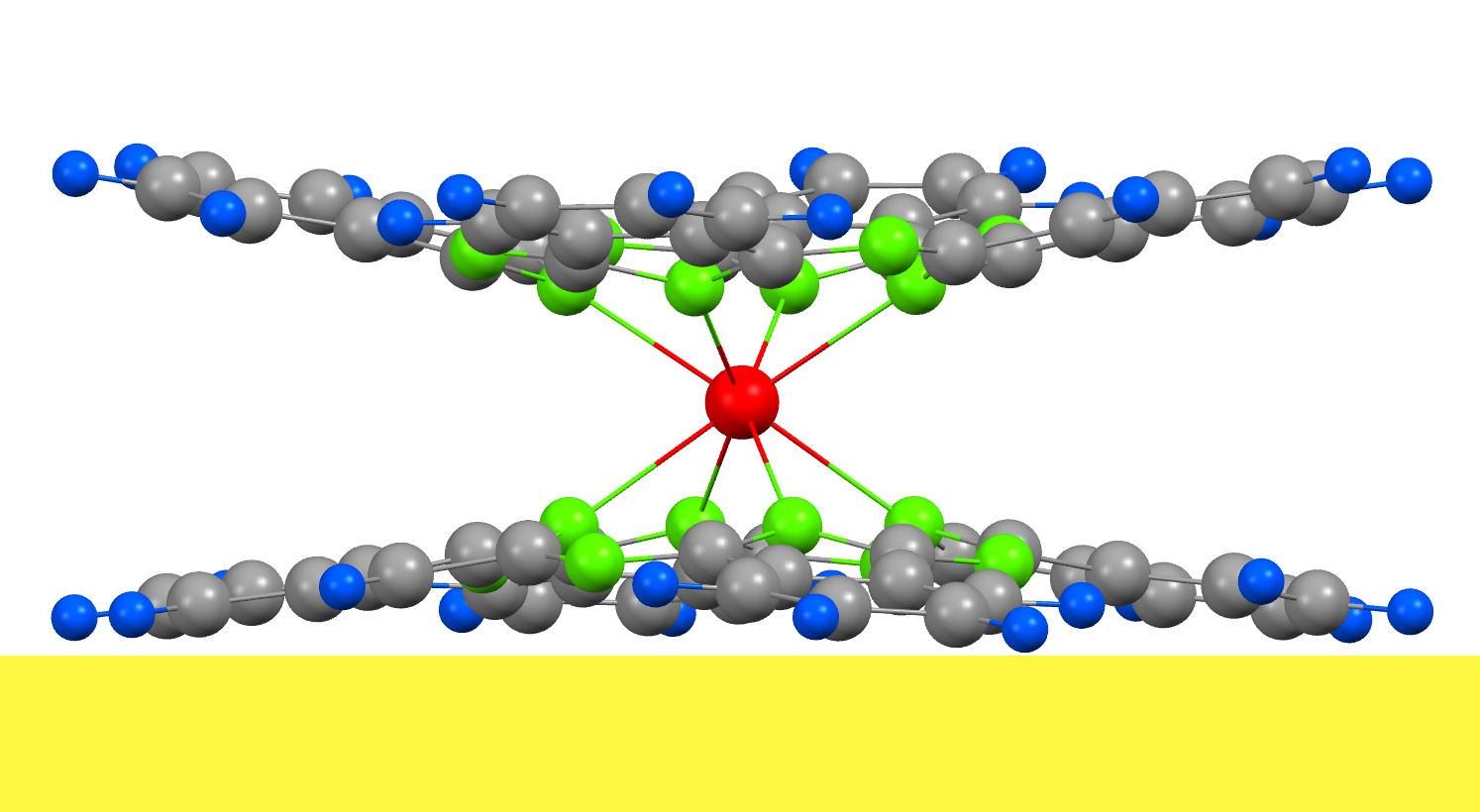}
\caption{TbPc$_2$ molecular structure formed by two organic ligands and a rare-earth atom of terbium, Tb, corresponding to the experimental geometry of Ref.~\onlinecite{Komeda2011}, in the configuration with a relative angle between the phthalocyanine complexes equal to 30$^\circ$. Red sphere marks terbium, Tb, while blue, green, and gray ones denote hydrogen (H), nitrogen (N), and carbon (C) accordingly. The molecule is deposited on a gold substrate (yellow slab).}
\label{genview}
\end{figure}

It has been suggested that the Kondo effect in TbPc$_2$ is caused by the spin of an unpaired electron occupying the aromatic $\pi$ orbital of the ligand. Importantly, one can change the relative position of top and bottom ligands (see Fig.~\ref{genview}) by application of external electrical pulses \cite{Komeda2011}, as was demonstrated by direct comparison of STM patterns for the top ligand with respect to crystallographic axes. Reshaping the ligand configuration has dramatic consequences for the Kondo effect: while for certain configurations the Kondo resonance in the differential conductance curve $dI/dV$ obtained in STM measurement is well pronounced (e.g., if the relative angle between the top phthalocyanine molecule and the bottom one is 45$^\circ$), it disappears for the others (e.g., if the relative angle between the top phthalocyanine molecule and the bottom one is 30$^\circ$). If Cu(111) rather than Au(111) is used as the substrate, the Kondo effect is absent for all ligand configurations \cite{Vitali2008}.

While experimental measurements clearly demonstrate that modifications in the electronic structure of the TbPc$_2$ molecule are mainly responsible for the tunability of the Kondo resonance, the corresponding microscopic mechanism remains, however, debated. In this paper, we present a theoretical analysis of the Kondo resonance and its tunability using {\em ab initio} theory. We make a direct comparison of our results with existing experimental data and analyze the presence and absence of Kondo features, in terms of charge transfer between the substrate and molecular magnet. Our calculations are based on a pseudopotential approach that unambiguously manifests the formation of the singly occupied molecular orbital (SOMO) for the geometries where the Kondo peak is detected, and the lack of the SOMO for those where the Kondo-peak is absent. To include the effect of the Au substrate, we determine the hybridization between the SOMO level of an SMM as obtained from {\it ab initio} calculations and plane waves that model conduction electrons of a metallic substrate. Note that the pseudopotential calculations are validated by a comparison to electronic structure theory within the projector augmented wave (PAW) method as implemented in VASP \cite{Kresse:94,Kresse:96} as well as a full-potential, all-electron method as implemented in RSPt \cite{RSPt}. The latter method also allows for an analysis of the nature of the $4f$ states, which is discussed below.

\section{Pseudopotential approach}

Although we attempt to make our analysis as general as possible, some assumptions have to be made for TbPc$_2$, in order to obtain quantitative information on what interactions and parameters most strongly affect the behavior of the SMM. It is well known that chemical structuring may be used to explore Kondo physics in a controllable fashion, e.g., by means of dehydrogenation \cite{Zhao2005}. However, in our simulations we assume that the chemical composition of SMM remains unchanged. In the following, we assume that TbPc$_2$ represents an isolated system, in the sense that the substrate does not influence the energy levels of the molecule significantly.

Detailed microscopic analysis of the experimental geometry of Ref.~\onlinecite{Komeda2011} requires a starting geometry optimization followed by density functional theory (DFT) calculations of both the equilibrium geometry and the electronic structure of TbPc$_2$. In our simulations, the preliminary geometry optimization of TbPc$_2$ complex is achieved by means of the semi-empirical approach labeled PM7 \cite{mopac16,maia12} implemented in the MOPAC2016 package. It is noteworthy that experimentally accessible structures with relative angles between the top and bottom ligands of 30$^\circ$ and 45$^\circ$ are found to be stable.

We further proceed with DFT simulations on the basis of the Firefly8 package \cite{firefly8}. It is well known, that the direct application of DFT-based methods to the systems with a partially occupied $f$ shell is questionable. Meanwhile, for the lanthanide series the $f$ shell is not involved in the formation of chemical bonding and is screened by itinerant valence bands electrons. We thus approximate the core electrons of the first four shells and eight $4f$ electrons with an effective pseudopotential, whereas the last $4f$ electrons of the Tb become bound with nitrogens. We exploit the ECP54MWB \cite{dolg89, dolg93} relativistic potential with the fixed $f$ subconfiguration and ($8s7p6d$)/[$6s5p5d$] valence basis set for trivalent Tb. Such configuration may be next utilized to fulfill self-consistency with other atoms characterized by ordinary non-relativistic $6-31G(d)$ basis sets. The accuracy of this computational scheme was tested against VASP calculations as well as all electron calculations, presented in Appendix~\ref{app:app}.

\section{Results}
\subsection{Electronic structure of TbPc\texorpdfstring{$_2$}{Lg} }

\begin{figure}[!h]
\includegraphics[scale=0.33]{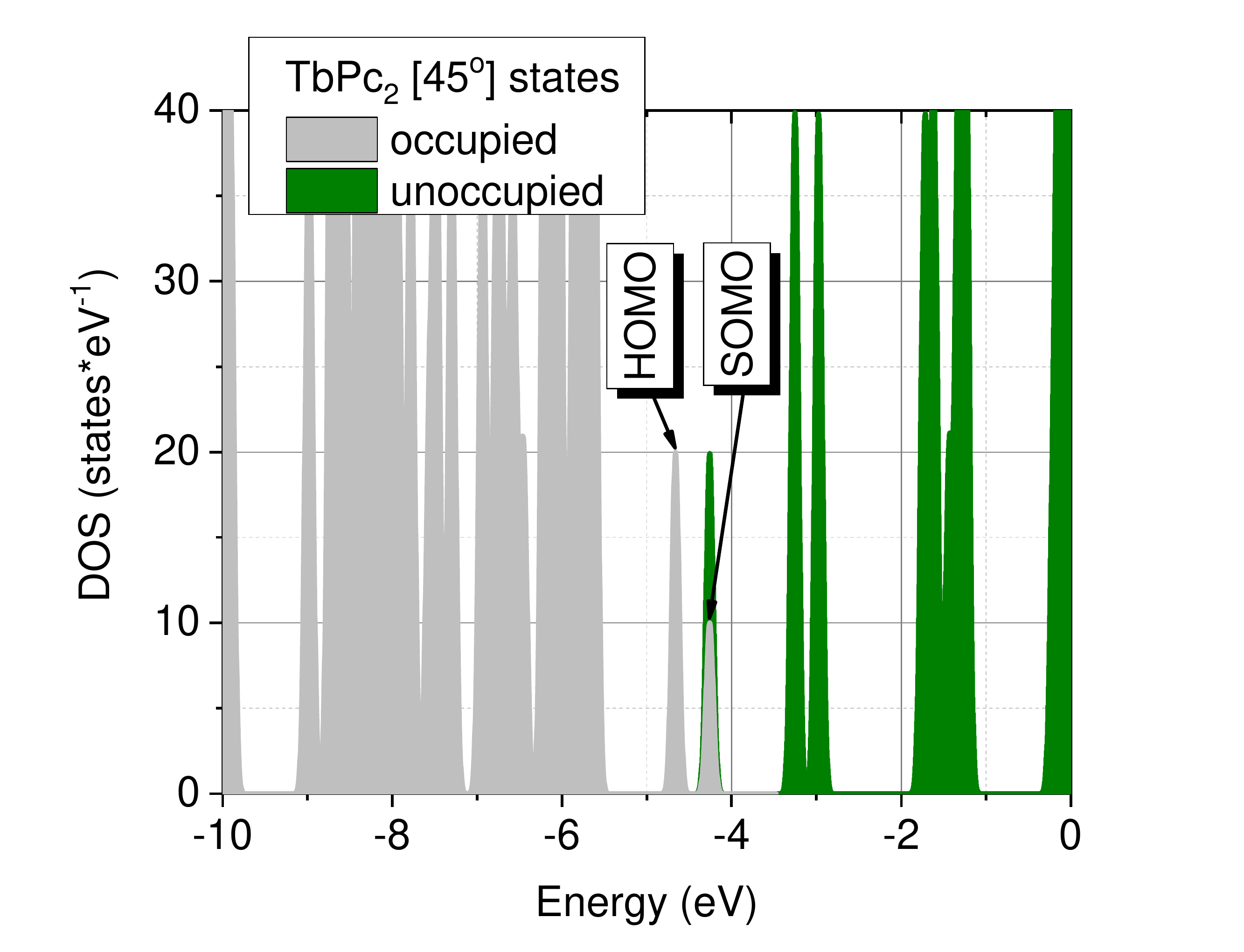}
\caption{Total DOS computed by the pseudopotential method. The gray filled curves show occupied states, and the green filled curves are for the unoccupied ones.}
\label{app1.1}
\end{figure}

With the geometry optimization performed in the first step of our calculations we determine self-consistently the atomic positions and the electronic structure in the form of density of states (DOS) for the 45$^\circ$-rotated Pc complexes as shown in Fig.~\ref{app1.1}. The results of our numerical calculations for a neutral molecule and singly ionized states are summarized in Table~\ref{OMOtable}, where eigenvalues of the SOMO and highest occupied molecular orbital (HOMO) are listed for the configurations with relative angle between the ligands equal to 30$^\circ$ or 45$^\circ$. Note that we have adopted the notation of Ref.~\cite{Komeda2011}. A neutral isolated molecule of TbPc$_2$ contains 593 electrons, an odd number of valence electrons, which implies that the SOMO must have at least one unpaired electron. This however does not mean that the Kondo effect is showing up. In fact,  to make that happen (i) the energy of the SOMO should be below the Fermi energy of the substrate to prevent the molecule from releasing an electron to the substrate with being switched to the positively charged nonmagnetic ion as a result; and (ii) the energy of the HOMO of the negatively charged ion has to be above the Fermi energy of the substrate. Note, that the direct use of DFT results for a comparison with experimental data, is unlikely to be reliable. Indeed, all existing exchange-correlation functionals yield systematic underestimation of the ionization potential \cite{zhang07}, and corresponding corrections have to be properly addressed. 

\begin{table}[]
\caption{Eigenvalues of the highest occupied orbitals of the neutral molecule as well as positively/negatively ($+/-$) charged TbPc$_2$, with the relative angle between ligands equal to 30$^{\circ}$ and 45$^{\circ}$. For positively charged molecules the energy of the lowest unoccupied molecular orbital is also given. All energies are measured in eV. For values listed in parentheses, see text for detailed description. An asterisk (*) marks the HOMO orbital that results from pairing up two electrons one of which belongs to the SOMO.}
\begin{tabular}{|p{2.1cm}||p{2.0cm} |p{2.0cm}|p{2.0cm}|}
\hline
Configuration   & calculated HOMO eigenvalues & calculated LUMO eigenvalues & calculated SOMO eigenvalues \\
\hline
\hline
TbPc$_2$[30$^{\circ}$]$^{+}$   &      $-7.09$  & $-6.73$ &   $ $ \\
TbPc$_2$[30$^{\circ}$]         &      $-4.55$  &   &     $-4.33$ ($-5.53$) \\
TbPc$_2$[30$^{\circ}$]$^{-}$   &      $-1.90^\ast$ &   &   $ $ \\
\hline
TbPc$_2$[45$^{\circ}$]$^{+}$   &      $-7.21$  & $-6.66$ &   $ $ \\
TbPc$_2$[45$^{\circ}$]         &      $-4.66$  &   &     $-4.26$ ($-5.46$) \\
TbPc$_2$[45$^{\circ}$]$^{-}$   &      $-1.84^\ast$ &   &   $ $ \\
\hline
\end{tabular}
\label{OMOtable}
\end{table}

DFT calculations reported here demonstrate that the energy of the SOMO for all possible configurations of a neutral molecule is above the Fermi energy of the Au substrate \cite{sachtler66,kahn16,Komeda2011}, as explicitly provided by the work function of gold at 5.3~eV below the vacuum energy with minor variations that concern substrate preparation and processing technology. For further details and an extensive list of experimental data and theoretical calculations of work functions see Ref.~\onlinecite{Skriver92}. In particular, the SOMO level is around $-4.3$~eV (see Table I), which is considerably higher than the estimated Fermi energy of $-5.3$~eV. Thus, the Kohn-Sham eigenvalue of the SOMO level (see Table~\ref{OMOtable}) has to be lowered down by about 1~eV for both orientations of the Pc complex to be consistent with the observed Kondo peak. Such an error, however, is not uncommon in these types of calculations \cite{zhang07}, primarily because the Kohn-Sham eigenvalue does not represent a proper quasiparticle energy. A self-energy correction is sometimes applied to the calculated Kohn-Sham energies \cite{kahn16} to remedy this error. 

Here we proceed with an analysis based on Slater's transition state theory. According to Janak's theorem \cite{Janak78} the Kohn-Sham eigenvalue, $\varepsilon_i$, of a specific state corresponds to the derivative of the total energy, $E_{tot}$, of the system with respect to the occupation, $n_i$, of that state, i.e., $\varepsilon_i=dE_{tot}/dn_i$ with the functional dependency $\varepsilon_i (n_i)$ in the most general case. This dependence can be quite strong, as exemplified by the results presented in Table I. The lowest unoccupied molecular orbital (LUMO) of the positively charged TbPc$_2$ is almost aligned with the SOMO state of the neutral molecule, with the difference being an occupation number $n_i=0$ in the former case and $n_i=1$ in the latter. The total energy required to remove one electron from the SOMO level can according to Janak's theorem be written as
\begin{eqnarray}\label{janaks}
\Delta E_{tot}=\int_0^1 \varepsilon_i (n_i) dn_i
\end{eqnarray}
For practical calculations of Eq.~\ref{janaks}, one would have to perform a series of Kohn-Sham calculations with various occupation numbers, $n_i$, and then do this integral numerically. However, the linear relation $\varepsilon_i\propto n_i$ typically holds in most of the cases, making only the calculation at $n_i=0.5$ sufficient to perform the integral, i.e., $\Delta E_{tot}=\varepsilon_i (0.5)$. Alternatively, and this is the choice we make here, one can replace $\varepsilon_i (0.5)$ with the average of its end points, i.e., 
$\Delta E_{tot}=(\varepsilon_i(1) + \varepsilon_i(0))/2$. This is the average of the SOMO energy of the neutral molecule and the LUMO energy of the positive charged molecule. Thus obtained values are listed in parentheses in Table~\ref{OMOtable}. We compare this estimate of the excitation energy of the single electron in the SOMO orbital with an analysis based on total energies of the neutral and the positively charged molecule. These results are summarized in Table~\ref{Janak}, and it may be seen that the energy difference in this table agrees well with the estimated SOMO excitation energy of Table~\ref{OMOtable}.

Interestingly, the energy of the SOMO level as listed in Tables~\ref{OMOtable} and ~\ref{Janak} is very close to the Fermi energy of Au for both geometries that have been investigated. Hence, one may conclude that theory gives more or less the correct energy of the SOMO level, in order to be consistent with the experimental detection of the Kondo effect. Unfortunately, the energy of the SOMO level of the molecule with 45$^\circ$ rotation is higher than that of the molecule with 30$^\circ$ rotation, albeit the difference between the two levels is tiny and approaches the accuracy of the calculations. 

\begin{table}[]
\caption{Total energies of neutral and positively charged TbPc$_2$, with the relative angle between ligands equal to 30$^{\circ}$ and 45$^{\circ}$}

\begin{tabular}{|p{2.1cm}||p{2.3cm} |p{2.1cm}|}
\hline
Configuration   & calculated $E_{tot}$ (hartrees) & $\Delta E_{tot}$ (eV) \\
\hline
\hline
TbPc$_2$[30$^{\circ}$]$^{+}$   &  $-3366.93$ & \\ \hline
TbPc$_2$[30$^{\circ}$]         &  $-3367.13$ & \\ \hline
                               &           & $-5.54$ \\ 
\hline
TbPc$_2$[45$^{\circ}$]$^{+}$   &  $-3366.94$ &\\ \hline
TbPc$_2$[45$^{\circ}$]         &  $-3367.14$ &\\ \hline
                               &           & $-5.47$\\
\hline
\end{tabular}
\label{Janak}
\end{table}

\subsection{Analysis of STM results}

In order to proceed with our analysis, we need to recall the relations between calculated microscopic properties and the observable experimental quantities. The first one is the origin of the low-voltage STM spectrum. The conductance can be estimated as \cite{bardeen61}

\begin{eqnarray}
\frac{dI(eV)}{dV} &=&\frac{4\pi e}{\hbar}\rho_{\rm tip} (E_F^{\rm tip})\rho_{\rm SMM} (E_F^{\rm SMM}+eV)\times \\ &\times& |M(E_F^{\rm SMM}+eV, E_F^{\rm tip})|^2,
\label{tuncur1}
\end{eqnarray}

\noindent where $\rho_{\rm tip}$, $E_F^{\rm tip}$ and $\rho_{\rm SMM}$, $E^{\rm SMM}_F$ are densities of states and Fermi energies for the tip and SMM correspondingly, $M$ is the tunneling matrix element, and $V$ is the bias voltage. Assuming that the temperature is low so that the density of unoccupied states can be considered as a step function, and the tunneling matrix element is independent of energy, it is possible to access the SMM density of states through measuring the $dI/dV$ curve. The second essential factor for the analysis is the brightness of the STM pattern for the given system measured with a fixed bias voltage. This phenomenological factor is derived as an integral over the $dI/dV$ spectrum and characterizes the overall density of states of the considered system up to the bias voltage.

Having basic knowledge of the electronic structure, one may interpret experimental data, which in the Kondo regime have a differential conductance with a three-peak structure (see supporting information Fig.~S4 in Ref.~\onlinecite{Komeda2011}). The zero-bias peak corresponds to the Kondo resonance at the Fermi level, the peak at $-0.1$~V is associated with a resonant tunneling through HOMO states, and the peak at $-0.9$~V is depicted but its origins are not discussed. Changing the relative angle of the molecule to 30$^\circ$ results in the experiments in a suppression of the zero-bias peak as compared to the case of 45$^\circ$.

Based on these considerations we propose three mechanisms of the Kondo resonance quenching: degradation of the hybridization between the SOMO level with the Fermi sea of the substrate and negative or positive charging. General consideration of the first and the second case complemented with our {\em ab initio} calculations demonstrates that they at least should not result in the decrease of the molecule density of states near the Fermi level, which is in contradiction with observed decrease of the STM brightness for the non-Kondo regime. Note that from our calculations this suppression may occur only due to the positive  charging of the SMM, resulting in the following model. When the molecule of a given, specific geometry hybridizes with the substrate the SOMO eigenenergy, for the case when the rotational angle is equal to 45$^\circ$, is slightly lower than the SOMO eigenenergy for the angle of 30$^\circ$. Therefore, the latter lies above the Fermi level of Au, and the former is slightly below. In the first case the HOMO and SOMO are both present and manifest themselves in the STM spectra and the SOMO creates the Kondo resonance. In the second case the molecule releases an electron into the substrate, charges positively, and fades on the STM pattern due to the shift of the density of state under bias voltage. Further, changes of the geometry will lead to the fine tuning of the energy of the SOMO in a controllable manner and corresponding tuning of Kondo effect, which was performed in Ref.~\onlinecite{Komeda2011} by applying the electrical pulses.

\subsection{Molecule-substrate exchange coupling}

Unfortunately, the level ordering in Tables I and II indicate that the SOMO energy is lowest for the molecule with 30$^\circ$ rotation of the Pc rings, which seems to be inconsistent with the fact that it is only for the 45$^\circ$ rotation of Pc rings the Kondo effect has been observed. One might from the experimental findings expect the SOMO energy of the 45$^\circ$ configuration to be just below the Fermi level of Au while the SOMO energy of the 30$^\circ$ configuration to be just above, which is at odds with the data in Tables I and II. It should be noted that the energy difference between the two molecular geometries is tiny and approaches a level where these kinds of calculations reach their limit of accuracy. However, in order to analyze whether the interaction between the molecular levels and the substrate may modify the situation, we present below an estimate of the energy change of the SOMO level due to the interaction with the substrate.

We start by assuming that a SMM is deposited on a substrate in the form of a slab occupying the lower half space ($z<0$) as shown in Fig.~\ref{genview}. We describe the states of conduction electrons, specified by the wavevector $\bm{k}=(\bm{k}_\parallel,\bm{k}_\perp)$ with the in-plane components $\bm{k}_\parallel$ and the out-of-plane component $\bm{k}_\perp=k_\perp\hat{\bm{e}}_z$ (here the unit vector $\hat{\bm{e}}_z$ is perpendicular to the substrate surface), inside the substrate ($z<0$), 
\begin{equation}\label{pwin}
	\Phi_{\bm{k}}(\bm{r})=\phi_{\bm{k}}(\bm{r})\left[\cos(\bm{k}_\perp\bm{r})-\frac{q_\perp}{k_\perp}\sin(\bm{k}_\perp\bm{r})\right],
\end{equation}
whereas being of evanescent nature outside the substrate ($z>0$),
\begin{equation}\label{pwout}
	\Phi_{\bm{k}}(\bm{r})=\phi_{\bm{k}}(\bm{r})e^{-\bm{q}_\perp\bm{r}},
\end{equation}
with the wave vector $\bm{q}_\perp=q_\perp\hat{\bm{e}}_z$ on condition that $q_\perp=\sqrt{2m_\ast(E_F+W)-k_\perp^2}$. The Fermi level of conduction electrons is positioned at $E_F$, while $W$ stands for the work function (energy needed to withdraw an electron from a metallic substrate). We have defined in Eqs.(\ref{pwin}) and (\ref{pwout}) a prefactor,
\begin{equation}
	\phi_{\bm{k}}(\bm{r})=\frac{e^{i\bm{k}_\parallel\bm{r}}}{\sqrt{V}}\left(\frac{k_\perp^2}{m_\ast(E_F+W)}\right)^{\frac{1}{2}},
\end{equation}
which characterizes the in-plane wave function of conduction electrons in the  volume $V$. It is noteworthy that the wave function of Eqs.~(\ref{pwin}) and (\ref{pwout}) is justified for parabolic dispersion of charge carriers with the effective mass $m_\ast$, something which is realistic for the energy levels that cut through the Fermi energy of Au. In our analysis, we keep the description of a magnetic molecule in terms of the SOMO wave function, $\Psi_\mathrm{SOMO}(\bm{r})$, obtained from {\it ab initio} calculations.

\begin{figure}[h!]
\includegraphics[scale=0.3,clip]{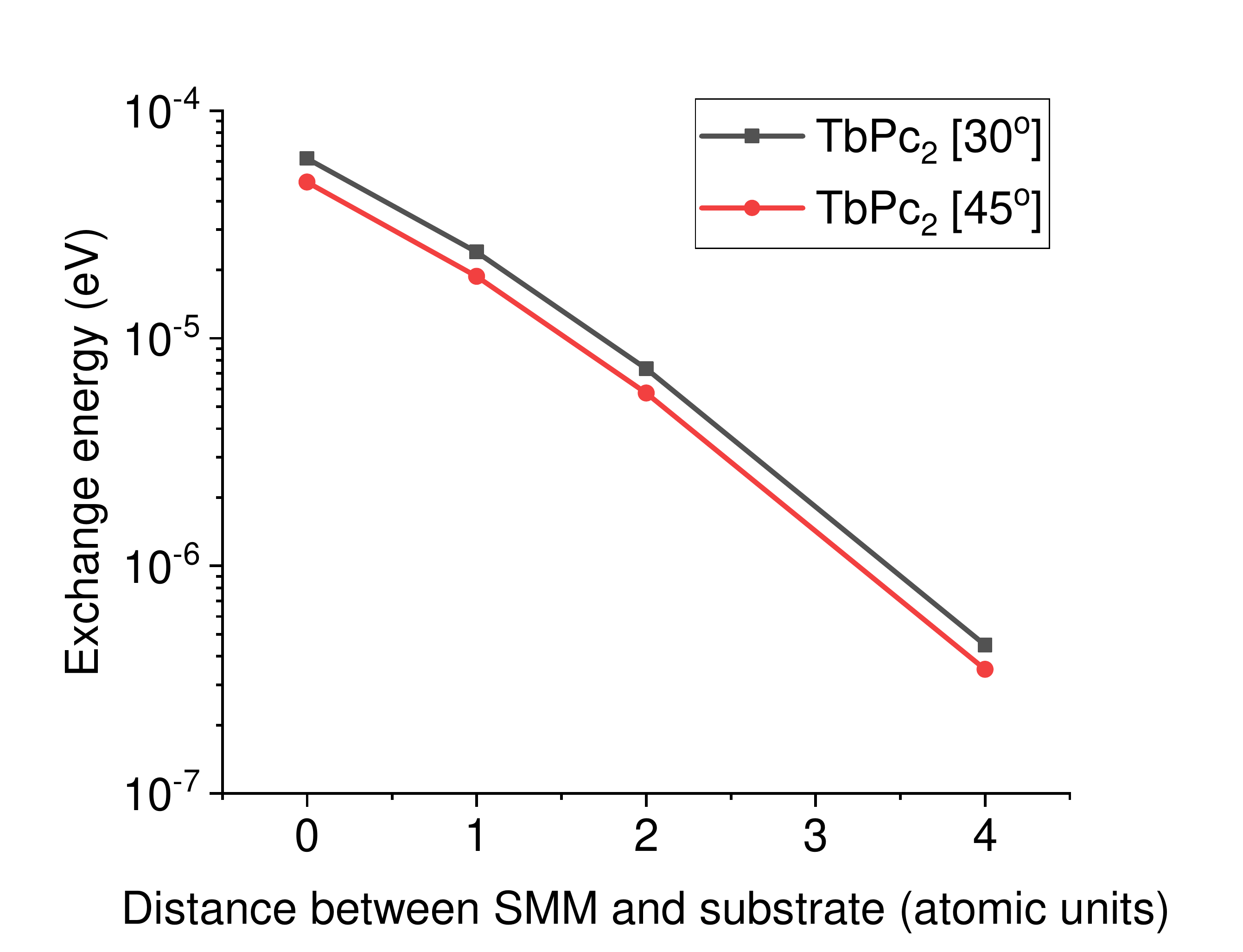}
\caption{Calculated dependence of the interaction energy between the SOMO level of TbPc$_2$ and substrate.}
\label{ehybdistdecay}
\end{figure}

Hybridization between the SOMO of the SMM and conduction electrons of the substrate can be quantified by evaluating the strength of the exchange interaction,
\begin{equation}\label{hyb}
	E_\mathrm{exch}=VN(E_F)\int\frac{d\Omega_{\bm{k}}}{4\pi}\mathcal{J}_{k_F}(\theta_{\bm{k}},\varphi_{\bm{k}}),
\end{equation}
where $N(E_F)$ is the number of states at the Fermi energy and averaging over solid angle $\Omega_{\bm{k}}$ is implied. The azimuthal, $\varphi_{\bm{k}}$, and the polar, $\theta_{\bm{k}}$, angles parametrize the Fermi surface, while the matrix element
\begin{align}\nonumber
	\mathcal{J}_{\bm{k}}=\int d\bm{r}_1\int d\bm{r}_2\Psi_\mathrm{SOMO}^\ast(\bm{r}_1)\Phi_{\bm{k}}^\ast(\bm{r}_2)\times \\ \label{ex}
	\times\frac{e^2}{|\bm{r}_1-\bm{r}_2|}\Phi_{\bm{k}}(\bm{r}_1)\Psi_\mathrm{SOMO}(\bm{r}_2),
\end{align}
specifies the strength of exchange coupling. For doing the integrals in Eq.~(\ref{ex}) we sample $\Psi_\mathrm{SOMO}(\bm{r})$ in the Cartesian frame; i.e., we define the wave function of the SOMO by its values $\Psi_\mathrm{SOMO}(\bm{r}_i)$ for a given set of points. The results of our numerical findings reveal that the two values $E_\mathrm{exch}=0.06$~meV for the configuration TbPc$_2$[30$^\circ$]  and $E_\mathrm{exch}=0.05$~meV for TbPc$_2$[45$^\circ$] are quite close to each other, with the tendency to be dramatically suppressed depending on the distance between SMM and substrate, as shown in Fig.~\ref{ehybdistdecay}. It is worth mentioning that the former results are obtained on the condition that the nucleus of the lowest atom of a magnet is brought in contact to the surface of a substrate (at $z=0$ plane); see Fig.~\ref{genview}. The dominant contribution to the integral of Eq.~(\ref{hyb}) stems from $\theta_{\bm{k}}=\pi/4$ and $3\pi/4$ and $\phi_{\bm{k}}$ such that the in-plane component of the wave vector $2\pi/\hbar|\bm{k}_\parallel|=15$ bohrs, i.e., exactly the diameter of the lobe of a benzene ring. The terms with $\theta_{\bm{k}}=0$ and $\pi$, corresponding to the product of the even wave function of the conduction electrons and odd wave function of the SOMO, as well as the ones selecting in-plane propagation $\theta_{\bm{k}}=\pi/2$ (and $\Phi_{\bm{k}}(\bm{r})=0$ as a result) do not contribute to Eq.~(\ref{hyb}). 

To summarize this subsection, the exchange interaction strength as estimated with Eq.~(\ref{hyb}) is on the order of a tenth of a meV for both TbPc$_2$[30$^\circ$] and TbPc$_2$[45$^\circ$] configurations, implying that recharging is a plausible mechanism for switching between the Kondo and non-Kondo regime. This is compatible with the results at a temperature of 10~K, which the authors of \cite{Komeda2011, Vitali2008} kept during their STM measurements, still being smaller as compared to detected Kondo temperature $T_K=31$~K. The results of numerical integration showing the sensitivity of the exchange energy to the distance between the SMM and substrate are shown in Fig.~\ref{ehybdistdecay} .

We note that a more rigorous theoretical analysis should also involve tiny experimental details, as well as the quality of substrate samples. In fact, as is seen from the expression in Eq.(\ref{hyb}) interaction energy is very sensitive to the number of particles at the Fermi level. This quantity at the given temperature drastically depends both on the band structure curvature at the Brillouin zone edges as well as the purity of the material. Particularly, for a substrate made of extremely pristine gold, e.g., with the concentration of impurities and imperfections less than 0.1\%, any metallic impurity serves as a donor center. To make this qualitative assessment one should notice that gold is characterized by almost the largest work function among all available materials; thus it can easily withdraw an electron from embedded impurities. Even for highly pristine samples of purity 99.95\% the number of such absorbed electrons, $N(E_F)$, at the Fermi level is two orders of magnitude larger as compared to the one due to thermal processes; a rough estimate provides a few meV from Eq.(\ref{hyb}) which can be translated to the Kondo temperature, $k_BT_K=3$~meV. 

Based on these arguments, one may speculate on the nature of the substrate for the Kondo peak to show up, in particular on the origin of inconsistency in the behavior of TbPc$_2$ on gold \cite{Komeda2011} and copper \cite{Vitali2008} substrate. Spectra from scanning tunneling spectroscopy unambiguously demonstrate that work functions for these two materials differ by 0.5~eV. The higher position of the Fermi energy in copper together with peculiar properties of its band structure pattern result in a lower probability of an impurity donating an electron to the substrate and a lower concentration of conduction electrons as a result. 

\section{Discussion and Conclusion}

In this paper, we have developed a systematic approach that describes Kondo-type features in the differential conductance of a magnetic TbPc$_2$ molecule deposited on a nonmagnetic Au(111) substrate. We have employed {\it ab initio} theory as a primary tool for our investigations, and focused on the energy levels of the molecular complex in relation to the Fermi level of the Au substrate. The transition of the neutral TbPc$_2$ molecule to the ionized state, TbPc$_2^+$, is suggested to be triggered by the change of the geometry of the organic ligands, which is induced by electric current pulses. We discovered that this mechanism is responsible for quenching of the Kondo resonance, by a careful analysis of the SOMO level as a function of molecular geometry. In contrast to the experimental results of Ref.~\onlinecite{Komeda2011}, we established that the energy of the SOMO level in the TbPc$_2$[45$^{\circ}$] configuration is higher than that of TbPc$_2$[30$^{\circ}$]. To address the effect of the substrate on the position of the SOMO level of the molecule we calculated the exchange interaction energy between the molecule and the substrate and showed that TbPc$_2$[30$^\circ$] and TbPc$_2$[45$^\circ$] isomers hybridize with the Au substrate in a similar manner. This emphasizes the relevance of previously made conclusions with regard to the nature of Kondo-regime switching and the importance of the bottom lobe arrangement with respect to substrate crystallographic plane \cite{Komeda2011}, because it is responsible for the hybridization with the Fermi sea. A certain discrepancy between experimentally determined Kondo temperature and our theoretical results on hybridization energy was also discussed.

\acknowledgements
A.A.P. acknowledges support from Russian Science Foundation Project No. 18-72-00058. O.E. acknowledges support from the Swedish Research Council (VR), the Knut and Alice Wallenberg Foundation (KAW), the Foundation for Strategic Research (SSF), eSSENCE, and STandUP. I.I.V. acknowledges support from Russian Foundation for Basic Research Project No. 20-52-S52001. The work of D.Y. was supported by the Swedish Research Council (Vetenskapsr{\aa}det, 2018-04383).

\appendix
\section{Benchmarking}\label{app:app}
To justify the use of the pseudopotential approach we perform DFT calculations using the VASP \cite{Kresse:94,Kresse:96} code and PAW method \cite{Kresse:99} to compute the geometric and electronic structure of TbPc$_2$ in the gas phase. The TbPc$_2$ molecule is inserted into a supercell of 25.9\,$\times$\,25.2\,$\times$\,16.7 {\AA} in order to minimize intermolecular interactions. The Perdew, Burke, and Ernzerhof (PBE) exchange correlation functional was used, the plane wave cutoff is 400 eV, and the long-range dispersion forces between the two Pc rings are incorporated by Grimme's second method (D2) \cite{Grimme:06}. The $4f$ orbitals of the Tb center are described by adding an effective Hubbard term $U_{\rm eff}$ of 5 eV, where the value of the Coulomb interaction parameter ($U$) was 5.7 eV and the exchange parameter ($J$) 0.7 eV \cite{Dudarev:98}. Spin-polarized calculations including spin-orbit coupling yield magnetic moments and structural geometry. For the DOS calculations presented in Fig.~\ref{app1.2}, a $\Gamma-$point-only $k$ mesh is employed.

\subsection{Density of states}
\begin{figure}[!h]
\includegraphics[scale=0.33,clip]{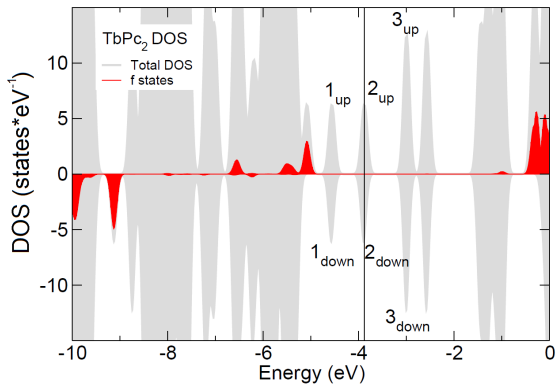}
\caption{Total DOS (gray filled curves) and $f$ DOS (red filled curves) calculated by DFT/PAW. Peak 1$_{\rm up}$/1$_{\rm down}$ is the HOMO, peak 2$_{\rm up}$/2$_{\rm down}$ is the SOMO, and peak 3$_{\rm up}$/3$_{\rm down}$ is the LUMO, in the spin up/spin down channels. The corresponding peaks in the lower part of the graph are the HOMO, SOMO, and LUMO in the spin down channel. A black line indicates the highest SOMO energy in the spin up channel at $-3.90$~eV, such that the DOS in the energy region above is unoccupied.}
\label{app1.2}
\end{figure}

Here we compare the pseudopotential DOS of the valence region, Fig.~\ref{app1.1}, with the total DOS and the $f$-DOS of TbPc$_2$ calculated by DFT/PAW, Fig.~\ref{app1.2}. In both cases we have considered a 45$^\circ$ ligand rotation as discussed in the paper. The DFT/PAW calculations give a magnetic moment of about $6\mu_B$. In the pseudopotential approach of Fig.~\ref{app1.1} the half occupied SOMO level is at $-4.2$~eV, the HOMO at $-4.6$~eV, and the LUMO at $-3.4$~eV. In the DFT/PAW calculations the HOMO is at $-4.58$~eV, the SOMO at $-3.90$~eV, and the LUMO at $-3.00$~eV. In the spin up states, the HOMO is at $-4.57$~eV, the SOMO at $-3.89$~eV, and the LUMO at $-3.00$~eV. Although the energy differences are slightly lower in the pseudopotential DOS, with a $\Delta E_\mathrm{HOMO-SOMO}=0.4$ eV as compared to the DFT/PAW $\Delta E_\mathrm{HOMO-SOMO}=0.7$~eV, it is evident that the main features of the total DOS in the valence region, both in the occupied and in the unoccupied parts, are the same in both approaches. From the DFT/PAW calculations we obtain that the $f$ state contributions to the DOS in the occupied part of the spin up channel are located below about 0.5 eV from the HOMO, and in the unoccupied part the first noticeable $f$ state contributions are seen about 0.9 eV above the LUMO. 

\subsection{A rigorous approach to 4\texorpdfstring{$f$}{Lg} subsystem}
\begin{figure}[h]
\includegraphics[scale=0.35,clip]{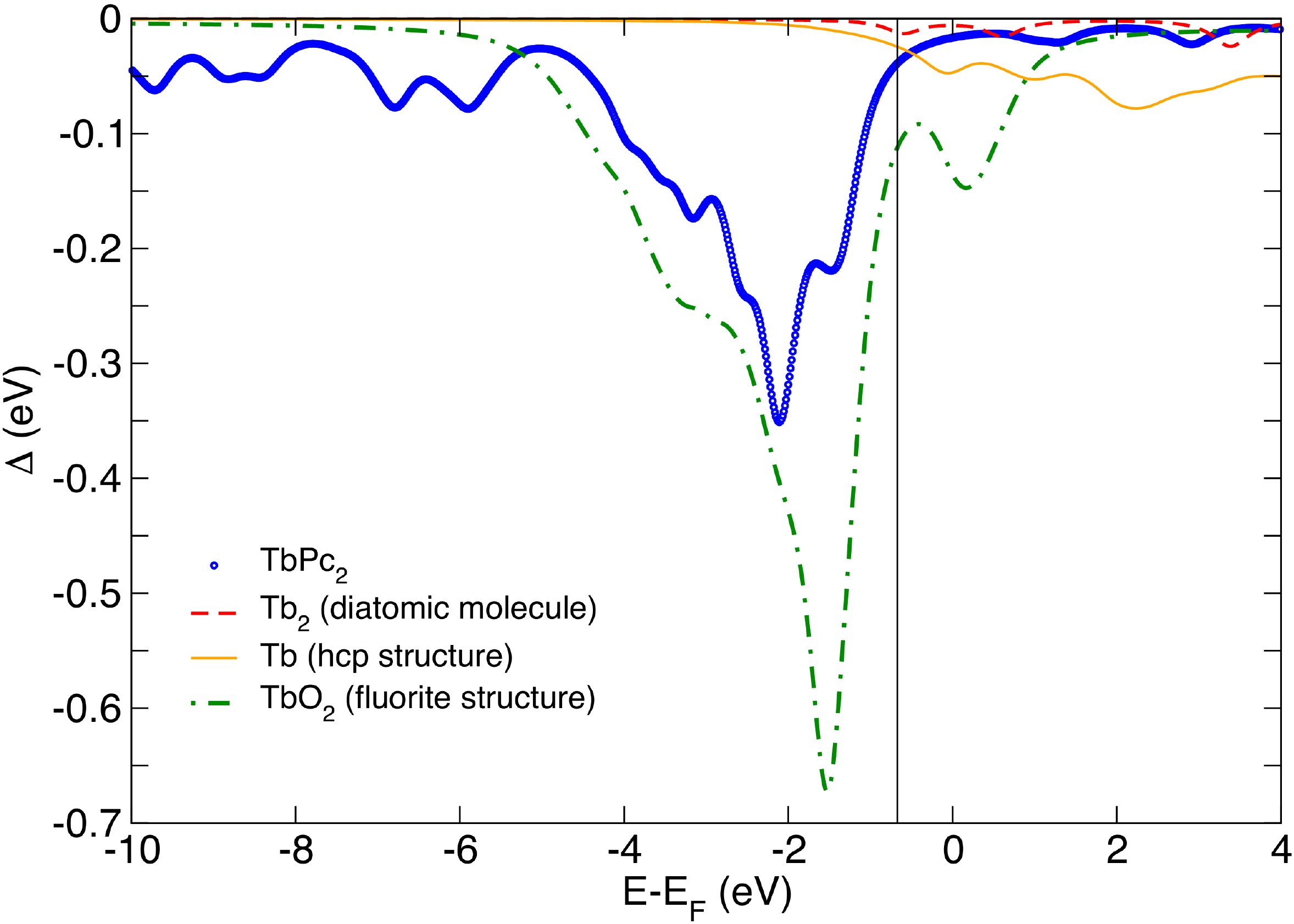}
\caption{Calculated hybridization function of $4f$ states with ligands of TbPc$_2$ (blue, dots), TbO$_2$ (green, dashed dotted line), Tb diatomic molecule (red, dashed), and bulk Tb (orange, full line). Bond distance of diatomic Tb is 3.17 \AA; for TbO$_2$ the lattice constant is 3.69 \AA; and for hcp, bulk Tb the hexagonal cell has lattice constants $a$ = 3.61 \AA\ and $c/a$ = 1.58. For TbPc$_2$ the nearest neighbor distance between Tb and N atoms is 2.43 \AA.}
\label{fig:hyb}
\end{figure}

The analysis presented so far assumes that the $4f$ level has a weak hybridization with the ligand N states. Such a model lies beyond the standard model of the rare-earth elements, which considers the 4$f$ shell on a single ion level. Therefore, the approach taken here  needs justification. One possible way to analyze the 4$f$ interaction with the other valence electrons is to look at the hybridization function, $\Delta(E)$. This allows us to quantify the degree of localization \cite{Herper:17}. In the framework of the Anderson impurity model, the hybridization function describes the interaction of a certain electron (in our case, Tb $4f$) with the surrounding bath of valence electrons. To understand the localization in TbPc$_2$ in detail the hybridization function of TbPc$_2$ has been calculated within the RSPt software \cite{RSPt}, together with some well known Tb systems, such as TbO$_2$, which serve as benchmark systems. Rare-earth dioxides are known for their quite strong interaction between the Tb $4f$ and the O $2p$ states and provide a large hybridization function which has in this case an expressed minimum around $1.52$~eV below the Fermi level, as shown in Fig.~\ref{fig:hyb}. Note that this is due to the crystal structure of the rare-earth dioxide series which is expected to give a very significant hybridization, as calculations for CeO$_2$ have demonstrated \cite{Herper:17}. The opposite behavior is expected from bulk Tb or a diatomic Tb molecule; both systems lack valence electrons other than from their own species and show a vanishing $\Delta(E)$ for occupied states (see Fig.~\ref{fig:hyb}). Comparing the calculated hybridization function of the TbPc$_2$ molecule to TbO$_2$ and the pure Tb systems (Tb$_2$, hcp Tb) shows that the hybridization function is surprisingly strong and the $4f$ electrons in the double-decker molecule are by far more itinerant compared to bulk Tb, since $\Delta(E)$ of the double-decker molecule is of the same order of magnitude as the one of TbO$_2$; see Fig.~\ref{fig:hyb}. The overall size of $\Delta(E)$ is slightly smaller than for TbO$_2$, being related to the fact that in comparison to oxygen, nitrogen offers one $p$ electron less to hybridize with. In addition, the distance between the Tb ion and the ligand plays a role. In the molecule the distance between the Tb and N$_{\rm iso}$ ions (2.43 {\AA}) is 7\% larger compared to the nearest neighbor Tb-O distance in TbO$_2$ (2.26 {\AA}). Note that the appearance of many small local minima in the hybridization function  which do not exist for the benchmark systems is due to the presence of the other ligands in the molecule.

Hence, the strong hybridization of $4f$ states with ligand orbitals (primarily next neighbor N$_{\rm iso}$ $2p$ states) of TbPc$_2$ suggests that they influence the electronic structure in a way that goes beyond the standard model of the rare-earth elements. This motivates the approach taken in the pseudopotential calculations with the Firefly8 package,  where we treated the 4$f$ shell as partly localized and partly itinerant. We end this section by noting that due to the lanthanide contraction, it is expected that lighter rare-earth ($RE$) elements in a series of $RE$Pc$_2$ molecules, would show even stronger hybridization, particularly CePc$_2$, the lightest rare-earth element with a significant filling of the $4f$ shell.

\bibliographystyle{apsrev4-1}
\bibliography{main.bbl}

\end{document}